\documentclass[
reprint,
superscriptaddress,
amsmath,
amssymb,
aps,
prb,
showpacs,
floatfix,
]{revtex4-2}

\usepackage{graphicx}% Include figure files
\usepackage{dcolumn}% Align table columns on decimal point
\usepackage{bm}% bold math
\usepackage{hyperref}% add hypertext capabilities
\begin{document}

\preprint{APS/123-QED}

\title{Rise and Fall of Anderson Localization by Lattice Vibrations:\\A Time-Dependent Machine Learning Approach}

\author{Yoel Zimmermann}
  \email{yzimmermann@ethz.ch}
 \affiliation{Department of Chemistry and Applied Biosciences, ETH Zurich, 8093 Zurich, Switzerland}
 \affiliation{Department of Physics, Harvard University, Cambridge, MA 02138, USA}

 \author{Joonas Keski-Rahkonen}
\affiliation{Department of Physics, Harvard University, Cambridge, MA 02138, USA}
\affiliation{Department of Chemistry and Chemical Biology,
Harvard University, Cambridge, MA 02138, USA}

 \author{Anton M. Graf}
\affiliation{Department of Chemistry and Chemical Biology,
Harvard University, Cambridge, MA 02138, USA}
\affiliation{Harvard John A. Paulson School of Engineering and Applied Sciences, Harvard University, Cambridge, MA 02138, USA}

 \author{Eric J. Heller}
   \email{eheller@fas.harvard.edu}
 \affiliation{Department of Physics, Harvard University, Cambridge, MA 02138, USA}
 \affiliation{Department of Chemistry and Chemical Biology,
Harvard University, Cambridge, MA 02138, USA}

\date{\today}

\begin{abstract}
The intricate relationship between electrons and the crystal lattice is a linchpin in 
condensed matter, traditionally described by the Fr{\"o}hlich model encompassing the 
lowest-order lattice-electron coupling. Recently developed quantum acoustics, 
emphasizing the wave nature of lattice vibrations, {has} enabled the exploration of 
previously uncharted territories of electron--lattice interaction not accessible with 
conventional tools such as perturbation theory. In this context, our agenda here is 
two-fold. First, we showcase the application of machine learning methods to categorize 
various interaction regimes within the subtle interplay of electrons and the dynamical 
lattice landscape. Second, we shed light on a nebulous region of electron dynamics 
identified by the machine learning approach and then attribute it to {transient 
localization}, where strong lattice vibrations result in a momentary Anderson prison for 
electronic wavepackets, which are later released by the evolution of the lattice. Overall, our research illuminates the spectrum of dynamics within the Fr{\"o}hlich model, such as 
transient localization, which has been suggested as a pivotal factor contributing to the 
mysteries surrounding strange metals. Furthermore, this paves the way for utilizing 
time-dependent perspectives in machine learning techniques for designing materials with 
tailored electron--lattice properties.
\end{abstract}

\maketitle

\section{Introduction}
Anderson localization refers to the cessation of diffusive wave propagation in disordered 
systems~\cite{Anderson_localization}. On~the historical front, Thouless theoretically 
posited~\cite{Thouless_Phys.Rev.Lett_39_1167_1977} that at low temperatures, where 
inelastic processes are minimal, localization would result in higher resistance compared to 
that expected from ordinary elastic scattering. This insight later spurred the development 
of the scaling theory of Anderson localization for non-interacting 
electrons~\cite{Abrahams_Phys.Rev.Lett_42_673_1979}. On~the other hand, the~conditions 
facilitating Anderson localization within an interacting system have been found to rely on 
several factors, including the strength of disorder, the~dimensionality of the 
system~\cite{Tarquini_Phys.Rev.B.95_094204_2017}, the~range and type of 
interactions~\cite{Gornyi_Phys.Rev.Lett_95_206603_2005, 
Fleishman_Phys.Rev.B_21_2366_1980, Fleishman_Phys.Rev.Lett_40_1340_1978}, and~the 
time scales of the disorder potential dynamics~\cite{Sacha_Phys.Rev.A_94_023633_2016, 
Phys.Rev.Lett_118_036602_2017}.

{The conundrum of whether systems localize or not} was recognized early on by researchers like Gogolin~\cite{gogolin1, gogolin2}, Thouless~\cite{Thouless_Phys.Rev.Lett_39_1167_1977}, and also Anderson~\cite{Anderson_localization, Phys.Rev_109_1492_1958}. For~instance, the~complex interplay between Anderson localization and lattice vibrations is observed in various random metal alloys and other disordered systems, such as crystalline organic semiconductors~\cite{Adv.Funct.Mater_26_2292_2016,Phys.Rev.Lett_96_086601_2006}  and halide perovskites~\cite{Phys.Rev.Lett_124_196601_2020}. The~random fluctuations caused by lattice motion gradually disrupt the quantum interference necessary for electronic state localization, leading to what has been coined {transient localization} (for capturing the essential aspects, see, e.g.,~Ref.~\cite{Phys.Rev.Lett_118_036602_2017}). This phenomenon combines aspects of both Anderson localized and itinerant electron systems: Electronic transport is characterized by the successive cycles of localization and delocalization à la Anderson stemming from lattice vibrations that eventually result in reduced~diffusion.

Whereas Anderson localization is typically explored within the framework of a tight-binding scheme featuring random on-site energies, the~standard model for lattice vibrations is established by Fr{\"o}hlich, which features linear coupling between an electron and the lattice. Conventionally, lattice vibrations are viewed through a number state perspective, but~the coherent state representation introduced in Ref.~\cite{Heller22}, known as {quantum acoustics}, treats lattice vibrations as waves rather than individual phonons.
This picture utilizing the coherent state basis is a valid way to treat the lattice vibrations fully quantum-mechanically, of~equal, unassailable stature to the conventional Fock (number) state approach. A~quantum lattice field
in a number state has a well-defined amplitude, i.e.,~the number of quanta, but~lacks knowledge of phase. On~the other hand, the~field defined by a coherent state has an 
equal amount of uncertainty in both amplitude and phase (a more detailed discussion on 
the coherent states can be found, e.g.,~in Refs.~\mbox{\cite{walls2007quantum, 
scully1997quantum, grynberg2010introduction, gerry2005introductory}}). However, even 
though these two pictures are equivalent at the most fundamental level, this duality is 
normally hidden by the approximations the two limits encourage. For~example, a~virtue of 
coherent states is that they are the closest quantum mechanical states to a classical 
description allowed by the uncertainty~principle.

The quantum-acoustical perspective unveils a duality between particle and wave pictures 
akin to quantum optics~\cite{walls2007quantum, scully1997quantum, 
grynberg2010introduction, gerry2005introductory}) established by 
Glauber~\cite{glauber1963}. Moreover, it allows for the electron--lattice interactions to be 
described in terms of a quasi-classical internal field, reminiscent of Bardeen and Shockley's 
concept regarding dynamical lattice distortions in nonpolar 
semiconductors~\cite{ShockleyBardeen1, ShockleyBardeen2}. In~particular, the~
deformation potential arising from lattice vibrations enables a quantum-coherent, 
nonperturbative treatment of charge carriers in coordinate space. In~addition to recovering 
the results of the conventional Bloch-Gr{\"u}neisen thory~\cite{Heller22}, the~program of 
quantum acoustics has illuminated mysteries surrounding strange metals where transient 
localization plays a central role, such as T-linear resistivity at the Planckian limit 
surpassing the Mott--Ioffe--Regel threshold~\cite{nature_manuscript} and a shift in the 
Drude peak in the optical conductivity towards the infrared 
range~\cite{DDP_manuscript}. Motivated by these advancements, we aim to identify 
various classes of dynamics hidden within the venerable Fr{\"o}hlich model, which we express in 
the coherent state~representation.

The quantum acoustical approach above enables the generation of large amounts of time-dependent charge carrier wavefunctions as a function of the system parameters. Clustering, a common unsupervised learning technique, provides an effective means to explore the spectrum of carrier behavior by grouping similar dynamical profiles into clusters. In~general, unsupervised machine learning (ML) methods have been established as a powerful tool to identify complex patterns in large unstructured data sets~\cite{alma99119072247205503, james2013introduction, carleo2019machine}. 

In the broader landscape of ML applications in physics, our approach aligns with the recent uses of machine learning to understand and categorize complex physical phenomena, such as many-body localization and phase transitions~\cite{carleo2019machine, zhang2019interpretable, van2017learning, carrasquilla2017machine, castro2023artificial, wang2016discovering}. However, it is important to distinguish our work from the common narrative of ``using ML to do physics''. Instead, our method uses ML as a tool that complements traditional analytical and numerical methods. This distinction underscores a shift from merely applying ML techniques to physics problems towards a more integrated approach where ML assists in how we conceptualize and explore physical~systems. 

To the authors' knowledge, this study is the first to apply ML techniques for analyzing the dynamics of condensed matter systems through a time-dependent lens. Moreover, our approach not only goes beyond the established focus on eigenstates but also extends the application of ML to condensed matter systems outside of tight-binding models, such as spin~chains.

Our program is as follows. In~Section~\ref{Sec:theory}, we delineate the theoretical 
framework across three stages. We first put forward the concept of deformation potential 
(Section~\ref{Sec:deformation_potential}), highlighting its significance as a 
palpable nonperturbative internal field for electrons (Section~\ref{Sec:electron_dynamics}). 
To facilitate the analysis of electron--lattice dynamics, we introduce a machine learning 
methodology in Section~\ref{Sec:clustering}. In~Section~\ref{Sec:results}, we present our 
classification of wavepacket dynamics leveraging the ML approach, exploring variations in 
the strength of the electron--lattice interaction and illustrating a resulting ``phase 
diagram''. Additionally, we conduct a detailed examination of one of the identified sectors 
connected to transient localization. Finally, we conclude our findings and discussions in 
Section~\ref{Sec:conclusion}.

%%%%%%%%%%%%%%%%%%%%%%%%%%%%%%%%%%%%%%%%%%%%%%%%%%%%%%%%%%%%%%%%%%%%%%%

\section{Theory and~Methods}\label{Sec:theory}

More explicitly, we investigate the diversity of physics contained by the following Hamiltonian:
\begin{equation}\label{Hamiltonian} 
\mathcal{H}_{\textrm{F}} = \sum_{\mathbf{p}}\varepsilon_{\mathbf{p}} c_{\mathbf{p}}c_\mathbf{p}^{\dagger} + \sum_{\mathbf{q}} \hbar \omega_{\mathbf{q}} a_{\mathbf{q}}^{\dagger} a_{\mathbf{q}} + \sum_{\mathbf{p}\mathbf{q}} g_{\mathbf{q}} c_{\mathbf{p} + \mathbf{q}}^{\dagger} c_{\mathbf{p}} \Big(a_{\mathbf{q}} + a_{\mathbf{-q}}^{\dagger} \Big)
\end{equation}
where $c_{\mathbf{p}}$ $(c_\mathbf{p}^{\dagger})$  is the creation (annihilation) 
operator for electrons with momentum $\mathbf{p}$ and energy 
$\varepsilon_{\mathbf{p}}$ whereas $a_{\mathbf{q}}$ $(a_\mathbf{q}^{\dagger})$ is the 
creation (annihilation) operator for longitudinal acoustic phonons of wave vector 
$\mathbf{q}$ and energy $\hbar \omega_{\mathbf{q}}$, respectively. The~
electron--phonon interaction is defined by its Fourier components $g_{\mathbf{q}}$. This 
Hamiltonian embodies the lattice $\mathbf{q}$, the~electrons $\mathbf{p}$, and~their 
lowest-order (linear) interaction that we next cast into the multimode coherent state basis 
of lattice degrees of freedom $\vert \chi \rangle$.

\subsection{Deformation~Potential}\label{Sec:deformation_potential}

The coherent state picture developed in Ref.~\cite{Heller22} is the dual partner of the 
traditional number state description of electron--lattice dynamics. In~this framework, each 
normal mode of lattice vibration with a wave vector $\mathbf{q}$ is associated with a 
coherent state $\vert \mathbf{q} \rangle$. At~thermal equilibrium, each mode can be 
considered to be equilibrated with a heat bath at temperature $T$, giving thermal 
ensembles of coherent states where the average occupation of the mode $\langle 
n_{\mathbf{q}}\rangle_{\textrm{th}}$ is given by the Bose--Einstein distribution. 
Employing the independence of normal modes, entire lattice vibrations can be described as 
the product state of the coherent states of the normal modes, in~other words, as a 
multimode coherent state $ \vert \chi \rangle = \bigotimes_{\mathbf{q}} \vert 
\mathbf{q} \rangle$, as~studied in Ref~\cite{heller_J.Phys.Chem.A_123_4379_2019}. 

Even though the Fock state perspective focusing on the particle characteristics of lattice 
vibrations and the coherent state viewpoint emphasizing the wave nature are formally 
equivalent, the~approximations they inspire are vastly different. For~instance, a~
common approach is to assume a direct product state $\vert \mathbf{p} \rangle \otimes 
\vert \chi \rangle$, combining the electronic state $\vert \mathbf{p} \rangle$ and the 
lattice state $\vert \chi \rangle$ while neglecting entanglement effects; this approach is 
equivalent to employing the time-dependent Hartree approximation.
Moreover, we only consider the longitudinal acoustic branch of lattice vibrations. Then, as~
detailed in Ref.~\cite{Heller22}, the~quasi-classical limit of quantum acoustics unveils a 
real-space, time-dependent description of electron--lattice interaction in terms of the 
\emph{deformation potential}.
\begin{align}
    V_D(\mathbf{r},t)
    = \langle \chi \vert  \sum_{\mathbf{q}} g_{\mathbf{q}} \Big(a_{\mathbf{q}} + a_{\mathbf{-q}}^{\dagger} \Big) \vert \chi \rangle \nonumber\\
    = \sum_{\substack{\mathbf{q}}}^{\vert \mathbf{q} \vert \le q_D}
    2g_{\mathbf{q}}
    \sqrt{\langle n_{\mathbf{q}}\rangle_{\textrm{th}}}
    \cos(\mathbf{q}\cdot\mathbf{r}-\omega_{\mathbf{q}}t+\varphi_{\mathbf{q}}),
    \label{eq:VDcl}
\end{align}
where $\varphi_{\mathbf{q}}$ is the phase of the coherent state $\vert \mathbf{q} 
\rangle$. Furthermore, we assume the phases $\varphi_{\mathbf{q}}$ to be uniformly 
distributed
random variables and employ the Debye model, assuming the linear dispersion 
$\omega_{\mathbf{q}} = v_s \vert \mathbf{q} \vert$, where $v_s$ is the speed of sound. 
Therefore, the~time dependence of the deformation potential is governed by the following 
wave equation:
\begin{equation}
    \frac{\partial^2}{\partial t^2} V_D(\mathbf{r},t) = v_s^2 \nabla^2 V_D(\mathbf{r},t).
\end{equation}

The acoustic lattice disorder field above appears as a chaotic sea of roaming sound waves, which can be loosely viewed as a dynamic, multi-wavelength adaptation of the Berry potential examined in Ref.~\cite{Kim_Phys.Rev.Lett_128_200402_2022}, named for its association with the random wave conjecture~\cite{Berry_J.Phys.A_10_2083_1977} in the field of quantum chaos. On the other hand, the deformation potential stemming from lattice vibrations has a close resemblance to the vector potential of a blackbody field as first identified by Hanbury Brown and Twiss~\cite{HBT}, except for the existence of the ultraviolet cutoff given by the Debye wavevector $q_D$ originating from the minimal lattice spacing $a$. 

The deformation potential  is a peculiar object. It is homogeneously random, meaning that the probability distribution of potential values $V_D$ does not depend on a position $\mathbf{r}$ or time $t$ (with the assumption of random phases). Therefore, each spatio-temporal patch of the potential is statistically indistinguishable from another. The typical length scale of the spatial correlations is determined by its largest wavenumber components $\sim q_D$. Similarly, the typical timescale of the potential change is determined by its largest frequency components $\sim \omega_D$. This special type of spatial-temporal correlation sets the deformation potential apart from other types of lattice distortions, as commonly investigated in the context of Anderson localization~\cite{Anderson_localization}.

\begin{figure*}[ht]
\includegraphics[width=0.9\linewidth]{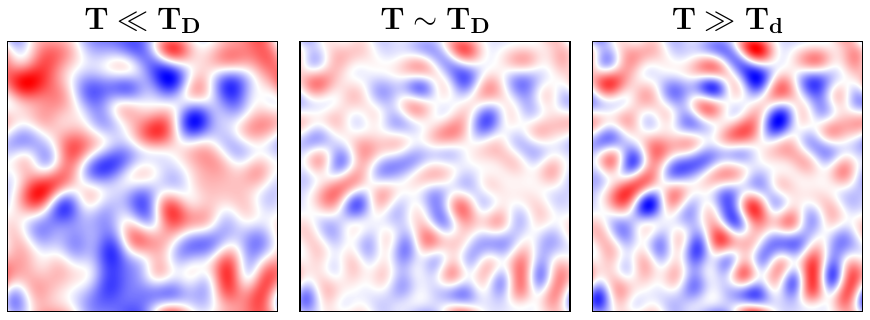}
\caption{Snapshots of the deformation potential at three different example temperatures. The~left and middle panels demonstrate the awakening of new vibrational modes with increasing temperature, giving rise to finer details in the potential. At~the same time, the~bumps {(red)} and dips {(blue)} of the potential become higher and deeper. On~the other hand, when the Debye temperature is reached, all the modes are active, and~the potential features grow as $\sim$$\sqrt{T}$, as~shown in the right panel. For~the sake of illustration, the~left panel has a different color scale than the middle and right~panels.}
\label{Fig:DP_temperature}
\end{figure*}

Even though the deformation potential overall averages to zero, its root-mean-square characterizing the strength of lattice disorder fluctuations, grows in temperature as 
\begin{eqnarray*}
 V_{\textrm{rms}}^{2} &=& \frac{2 E_d^2 \hbar}{\pi \rho v_s} \int_0^{q_D} \frac{q^2\textrm{d} q}{e^{\hbar v_s q/k_BT}-1} \nonumber\\ &\sim&
 \begin{cases}
(k_BT)^{1/2}, \quad \textrm{when} \quad T \gg T_D\\
(k_BT)^{3/2}, \quad \textrm{when} \quad T \ll T_D
\end{cases},
\end{eqnarray*}
where $E_d$ is the deformation potential constant related to the coupling $g_{\mathbf{q}}$, and $\rho$ is the mass density of the underlying crystal lattice. As the temperature nears the Debye temperature $T_d$, previously dormant vibrational modes start to awaken from their Bose-Einstein slumber. This activation not only enhances the peaks and valleys as $\sim T^{3/2}$ but also brings forth finer wavelength details in the deformation potential, as depicted in the left and middle panels of Fig.~\ref{Fig:DP_temperature}. At a temperature $T\sim T_D$, all the possible lattice modes are in play, after which no new wave characteristics emerge. The existing potential bumps and dips simply become more pronounced as $\sim \sqrt{T}$, as illustrated via the middle and right panels of Fig.~\ref{Fig:DP_temperature}.

\subsection{Electron~Dynamics}\label{Sec:electron_dynamics}

The time-varying deformation potential virtually demands quantum wavepacket propagation techniques for the electron. Here, we focus on the time-dependent Hamiltonian
\begin{eqnarray*}
    \mathcal{H}_0 = \frac{\vert \mathbf{p} \vert^2}{2 m^*} + V_{D}(\mathbf{r}, t),
\end{eqnarray*}
where $m^*$ is the effective (band) mass of the electron and $ V_{D}(\mathbf{r}, t)$ is the 
deformation potential given by Equation~\eqref{eq:VDcl}. This effective Hamiltonian 
$\mathcal{H}_0$ represents the electron component of the Fr{\"o}hlich model defined 
previously in Equation~\eqref{Hamiltonian} within the framework of the effective mass 
approximation.

Our investigation of electron dynamics under the defined effective Hamilton 
$\mathcal{H}_0$ approaches the issue from the point of view of Gaussian wavepackets 
that are a common tool for analyzing time-dependent aspects of a quantum 
system~\cite{heller2018semiclassical, tannor2007introduction}, for~instance in the studies 
of quantum optics~\cite{scully1997quantum, walls2007quantum}, 
scarring~\cite{keski2019quantum, PhysRevB.96.094204.2017, keski2024antiscarring}, and~
branched flow~\cite{heller2021branched, superwire1, superwire2}. Here, we choose the 
following test Gaussian for representing the charge carrier:
\begin{equation}\label{Initial_Gaussian}
    \Psi(\mathbf{r}, 0) = \mathcal{N}\exp\left( \frac{1}{4} \vert \mathbf{r} \cdot \boldsymbol{\sigma} \vert^2  - i \mathbf{k} \cdot \mathbf{r} \right),
\end{equation}
where $\mathcal{N}$ is the normalization factor, and~$\boldsymbol{\sigma} = 
(\sigma_x^{-1}, \sigma_y^{-1})$ describes the initial width of the wavepacket. Without~
loss of generality, we can choose to launch the test wavepacket into the x direction with 
the Fermi momentum, thus $\mathbf{k} = (k_F, 0)$, where $k_F$ is the Fermi wavevector. 
The memory of the initial form of the wavepacket is quickly lost in the chaotic potential 
and its exact form is~unimportant.

\begin{figure*}[ht]
\centering
\includegraphics[width=0.95\linewidth]{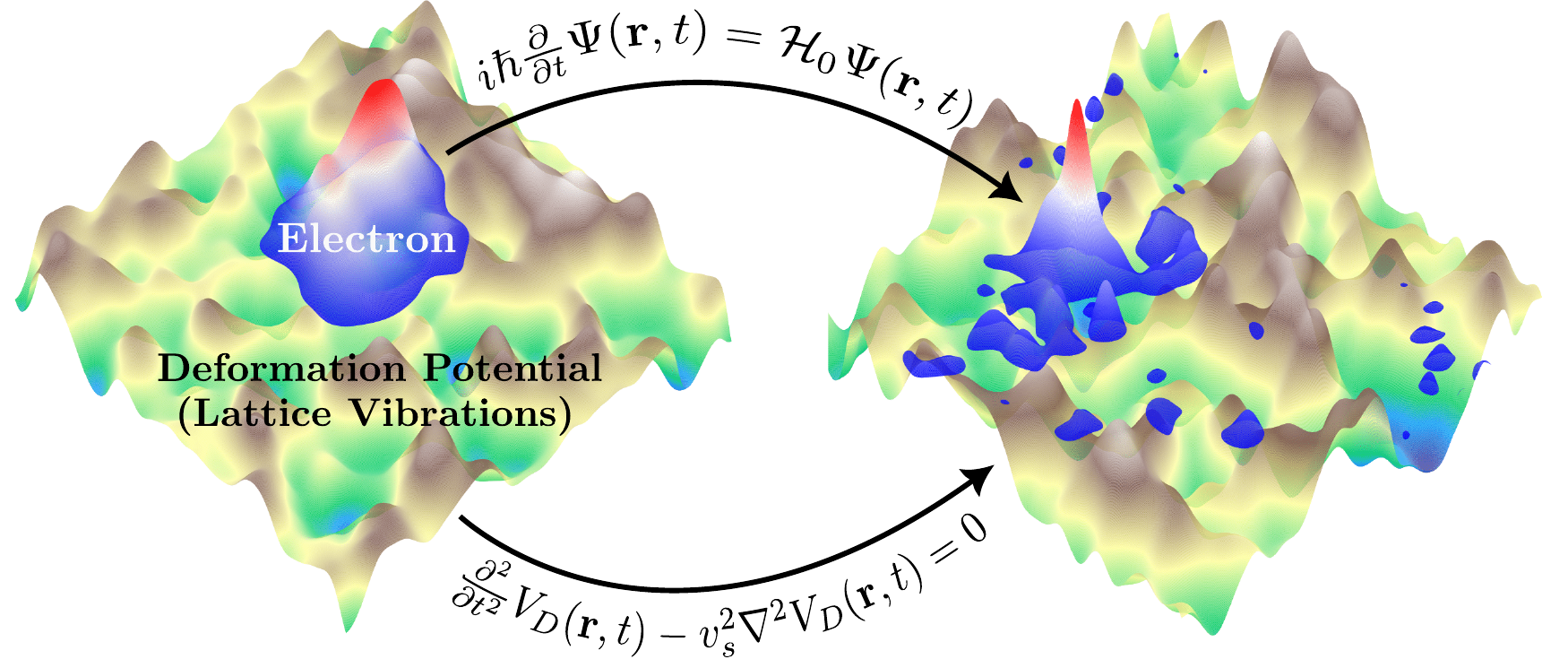}
\caption{The quantum acoustical Wave-on-Wave (WoW) approach to charge carrier dynamics. An~electron wavepacket propagates atop a deformation potential, which itself 
evolves according to the wave equation. As~it traverses this shifting acoustic landscape 
shaped by acoustic deformations, the~electron undergoes quasi-elastic scattering akin to 
impurity scattering.\label{fig:wave-on-wave}}
\end{figure*}

To propagate the wavepacket in time, we utilize the third-order split operator\linebreak method \cite{SplitOperator, Feit1982, tannor2007introduction, heller2018semiclassical} applied to the time-dependent Schr{\"o}dinger equation:
\begin{eqnarray}
    i \hbar \frac{\partial}{\partial t} \Psi(\mathbf{r}, t)=\mathcal{H}_0 \Psi(\mathbf{r}, t).
\end{eqnarray}
Figure~\ref{fig:wave-on-wave} illustrates the charge carrier wave, originally a Gaussian 
as described in \mbox{Equation~\eqref{Initial_Gaussian}}, evolving under the influence of the dynamic 
lattice wave field, which converts the always-accessible wave nature of lattice vibrations into 
something valuable, a~point where the quantum-acoustical perspective becomes tangible. 
Within this kind of Wave-on-Wave (WoW) approach, as~detailed in 
Refs.~\cite{DDP_manuscript, nature_manuscript}, one winds up solving two interacting 
equations of motions simultaneously: one for the lattice and one for the electron (i.e.,~the 
time-dependent Schr{\"o}dinger equation for the charge carrier and the wave equation for 
the lattice vibrations). 

In general, we can crudely characterize the WoW dynamics based on two criteria:
\begin{align*}
\bar{K} = \frac{\hbar^2 k_F^2}{2 m^* V_{\textrm{rms}}}
 \begin{cases}
	\gg 1  & \rightarrow \;  \text{Perturbative} \\
	\lesssim 1 \;  & \rightarrow \;  \text{Nonperturbative},
\end{cases}\end{align*}
and
\begin{align*}
\bar{\lambda} = \frac{\pi}{k_Fa} 
\begin{cases}
	\ll 1  & \rightarrow \;  \text{Incoherent} \\
	\lesssim 1 & \rightarrow \;  \text{Coherent}.
\end{cases}
\end{align*}

Here, we want to point out that the term ``coherence'' is reserved
to describe the spatial phase coherence of the electron wavefunction, not to be conflated 
with the ``coherent versus incoherent metals'' nomenclature, which pertains to the 
breakdown of the quasi-particle paradigm. Rather, this criterion refers to the quantum 
coherence of electrons that becomes important in scattering when the wavelength of the 
electrons (Fermi wavelength) is not much less than twice the lattice constant $a$. The~
WoW approach presented here adeptly captures the persistence of coherence between 
successive collisions, a~facet commonly overlooked by conventional Boltzmann transport 
methods. Indeed, the~preservation of coherence beyond the first scattering event can wield 
significant influence, as~evidenced by Refs.~\cite{Heller22, nature_manuscript}.

On the other hand, the~comparison of the kinetic energy of the electron (a fair 
approximation is given by the Fermi energy) with the root mean square of the deformation 
potential determines whether the lattice vibration and its resultant electron scattering can 
be treated perturbatively or not. In~essence, the~deformation potential cannot be merely 
considered a minor perturbation to the free-electron model Hamiltonian.
Instead, it can result in a substantial effect on the electronic density of states, as~shown in Refs.~\cite{Heller22,Phys.Rev.B_107_224311_2023}.

\subsection{Clustering}\label{Sec:clustering}

In addition to this classification based on the static properties of the electron--lattice 
interaction, we here explore the dynamical aspects of this relationship by examining two 
distinct measures, namely the {mean squared displacement}
(MSD) and 
{inverse participation ratio} (IPR). The~spread of the wavepacket over time is 
measured by the MSD:
\begin{eqnarray}\label{eq:MSD}
    \alpha(t) = \int \Psi^*(\mathbf{r}, t) \left[ \langle \mathbf{r} \rangle - \mathbf{r} \right]^2 \Psi(\mathbf{r}, t)\, \textrm{d}\mathbf{r}.
\end{eqnarray}
Moreover, we assess the level of wavepacket localization by considering the IPR
\begin{eqnarray} \label{eq:IPR}
    \beta(t) = \int \vert \Psi(\mathbf{r}, t) \vert^4\, \textrm{d}\mathbf{r},
\end{eqnarray}
 a widely used method for analyzing scarred states or Anderson localized states in a 
 disordered medium~\cite{Thouless_1974}. Here, it is important to note that the measures 
 discussed here deviate from their conventional definition by being determined as a 
 function of time rather than time-independent, as~typically seen in the studies of 
 eigenfunctions. Furthermore, we can combine the time-evolving quantities into one 
 two-dimensional times series, denoted~as 
\begin{eqnarray*}
    \mathbf{F}(t) = \begin{pmatrix} \alpha(t) \\ \beta(t) \end{pmatrix} \; ,
\end{eqnarray*}
{which}
paves the way to leveraging ML methods for time series to discern various 
transport~regimes.

Specifically, we apply $k$-means~\cite{Lloyd1982k-means} clustering using dynamic time 
warping~\cite{sakoe1978dynamic} to the mean-variance normalized set of series 
$\{\mathbf{F}(t)\}$, which consist of 50 timesteps with each timestep representing 
$2\,\textrm{fs}$ of evolution, across different system variable settings. At~its core, 
$k$-means is an algorithm to solve the optimization problem of partitioning a given set 
into $k$ clusters such that the in-cluster variance is minimized (for a detailed explanation 
of $k$-means and our method, we refer the reader to Appendix 
\ref{Appendix:Clustering}). It is important to stress that this optimization is performed in 
a 
fully {unsupervised} manner, meaning that it only processes the raw time series 
$\{\mathbf{F}(t)\}$ and is blind to the system variables used to generate the data. To~
ensure robustness and to mitigate the effect of statistical fluctuations, we average the 
clustering results of an ensemble of 10 time series data sets, each generated using 
randomly initialized deformation potentials. This method allows us to objectively identify 
unique clusters corresponding to diverse dynamical regimes hidden within the 
Fr{\"o}hlich~Hamiltonian.
%%%%%%%%%%%%%%%%%%%%%%%%%%%%%%%%%%%%%%%%%%%%%%%%%%%%%%%%%%%%%%%%%%%%%%%

\section{Results}\label{Sec:results}

As a real-world example of electron--lattice dynamics, at~least within the  Fr{\"o}hlich 
Hamiltonian, we investigate the prototypical strange metal {lanthanum strontium 
copper oxide} (LSCO), renowned for its diverse physics~\cite{Varma2020}.  This canonical 
cuprate, discovered by Bednorz and M\"uller~\cite{bednorz1986possible}, is characterized 
by an orthorhombic space group~\cite{Radaelli_phys.rev.b_49_4163_1994}. The~main 
electrical transport occurs between the $\textrm{CuO}_2$ layers, making it effectively 
two-dimensional in nature~\cite{Varma2020}. Furthermore, LSCO has a large 
electron--lattice coupling, and~its Fermi energy is adjustable through doping. The~material 
parameters for optimal doping are given in Table~\ref{tab:material} based on 
experimental values derived from Refs.~\cite{Padilla_Phys.Rev.B_72_060511_2005, 
Walsh_Phys.Rev.B_106_235134_2022, Bozovic_Phys.Rev.Lett_89_107001_2002, 
Legros_Nat.Phys_15_142_2019, Fang_Nat.Phys_18_558_2022}. These serve as the basis for 
constructing an associated deformation potential. These parameters align with previous 
investigations on strange metals~\cite{DDP_manuscript, nature_manuscript} using a 
quantum-acoustical perspective. Here, we explore electron--lattice dynamics in a broader 
scope, rather than focusing on specific attributes like electrical~conductivity. 

To enable this analysis, we introduce two scaled variables that we vary in our simulations: 
dimensionless temperature $ \Tilde{T} = T/T_D$ and effective electron--lattice coupling 
$\tilde{G} = 2k_F/q_D$. The~coupling is adjusted by varying the Fermi wavevector $k_F$ 
(energy) of the electron while maintaining the underlying lattice structure constant (with a 
fixed Debye wavevector $q_D$). This ensures that our variables $\tilde{T}$ and 
$\tilde{G}$ are independent, a~premise supported by the evidence from our~simulations.  

\begin{table*}[ht]
\caption{\label{tab:material}%
Material parameters for LSCO that are used in the simulations for constructing the deformation potential. 
}
\begin{ruledtabular}
\begin{tabular}{ccccccccc}
\textrm{Parameter} & $n$ [$10^{27} \mathrm{~m}^{-3}$] & $m^*$ [($m_e$)] & $v_s$ [$\mathrm{m} / \mathrm{s}$] & $E_d$ [$\mathrm{eV}$] & $\rho$ [$10^{-6} \mathrm{~kg} / \mathrm{m}^2$] & $E_F$ [$\mathrm{eV}$] & $a$ [\AA] & $T_D$ [$\mathrm{K}$] \\
\colrule
\textrm{LSCO} & 7.8 & 9.8 & 6000 & 20 & 3.6 & 0.12 & 3.8 & 379 \\
\end{tabular}
\end{ruledtabular}
\end{table*}

\subsection{Phase~Diagram}

\begin{figure}[h!]
\includegraphics[width=\linewidth]{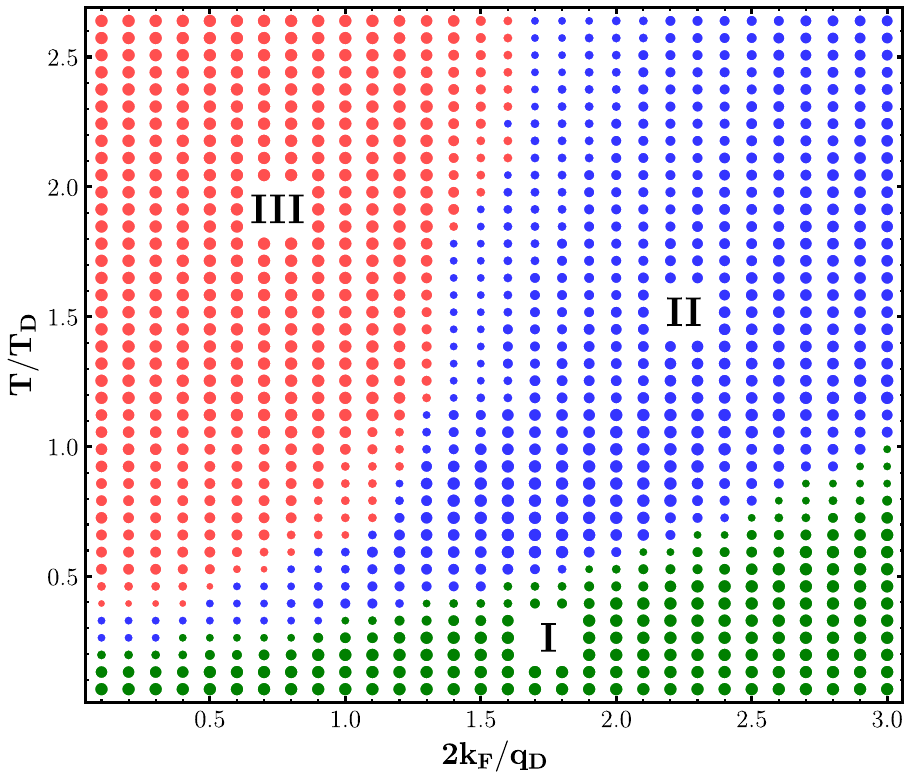}
\caption{Phase diagram of LSCO in the dynamic potential field. The~phase diagram was derived using a machine learning-based clustering algorithm to analyze time series data of the wavefunction evolution within systematically varied deformation potentials. This analysis involved variations in temperature $\tilde{T}$ and effective coupling $\tilde{G}$; every point corresponds to one unique configuration. The~following three clusters were identified: (I) refractive scattering region (green), (II) diffraction behavior (blue), and (III) short-time localization at high temperatures (red). The~size of the points indicates the level of agreement across an ensemble of different wavefunction data~sets.\label{fig:clusters_dynamic}}
\end{figure}

Our central finding is presented in Figure~\ref{fig:clusters_dynamic} showing the 
dynamical data classified using the ML-based clustering algorithm explained above and as 
detailed in Appendix~\ref{Appendix:Clustering} as the temperature $\Tilde{T}$ and 
coupling strength $\Tilde{G}$ are varied. Three distinct phases are identified as labeled by 
the differently colored regions. We want to emphasize that the term ``phase'' is used here 
to refer to regimes of different dynamical behavior, not in the thermodynamical sense. 
There are no sharp boundaries between these phases; the changes are gradual rather than 
true phase transitions. This fact is highlighted in Figure~\ref{fig:clusters_dynamic} by the 
different sizes of the points, representing the level of agreement within the ensemble of 
studied wavepackets for the given~parameters. 

\begin{figure}[h!]
\includegraphics[width=\linewidth]{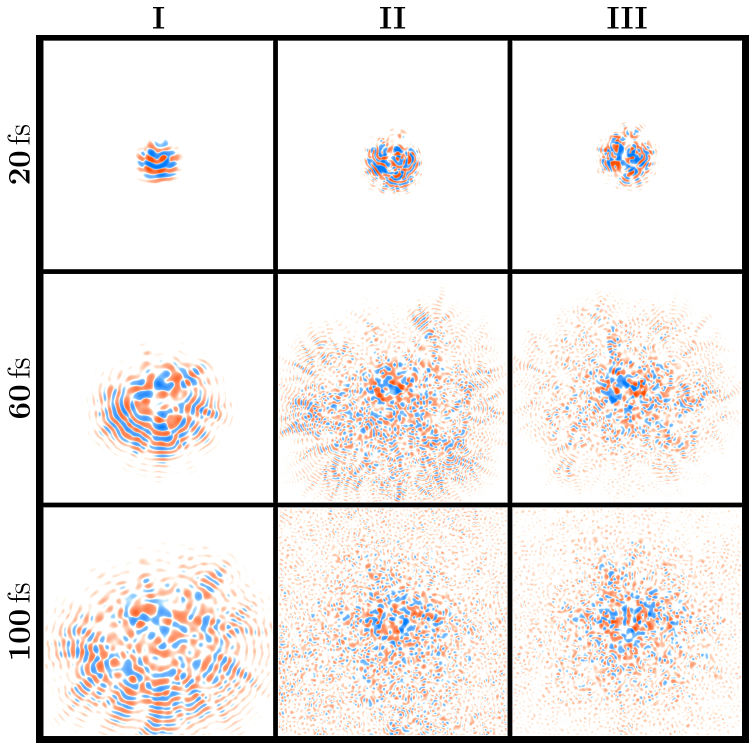}
\caption{Evolution of electronic wavefunctions in dynamic deformation potentials. This chart displays the real part of the wavefunction in two-dimensional coordinate space, where red and blue colors indicate positive and negative amplitudes, respectively. Each column represents parameters selected as examples from the identified clusters I, II, and~III. For~each cluster, the~panels arranged vertically from top to bottom show snapshots of the wavefunction at increasing times of 20 fs, 60 fs, and~100~fs.  \label{fig:wavefunc_dynamic}}
\end{figure}

We interpret the distinct regions as follows: refractive scattering phase (I), diffraction behavior phase (II), and~transient localization phase (III). We present three zones of characteristic wavepacket evolution, selected to represent the dynamical behavior of each phase. The~snapshots in Figure~\ref{fig:wavefunc_dynamic} depicts the real part of the evolution of a common initial Gaussian wavepacket at times of $20\,\textrm{fs}$, $60\,\textrm{fs}$ and $100\,\textrm{fs}$ under three different conditions of temperature $\tilde{T}$ and coupling strength $\tilde{G}$.

Phase I (green) is characterized by an almost linear phase boundary starting at\linebreak $\Tilde{T} \approx 0.45$ rising across the range of $\tilde{G}$ explored. This phase is perturbative in the sense of $\tilde{K} \gg 1$. As~seen in the left column of Figure~\ref{fig:wavefunc_dynamic}, the~scattering of the wavepacket is mainly refractive. This trend will lead to branched flow behavior~\cite{heller2021branched} at longer times; a propagating wave forms tree-like branches under a weakly disordered medium, due to small-angle refraction~\cite{topinka2001coherent}. Moreover, there is a partial transparency of the electrons to any shorter wavelength modes ($q > 2k_F$) present in the underlying deformation potential, as~is further explained in Ref.~\cite{Heller22}.

Phase II (blue) covers the upper right section of the phase diagram with high values of 
temperature variability ($\Tilde{T}$) and effective coupling($\tilde{G}$) and is separated by an exponential-like phase boundary from Phase III (red), which is characterized 
by high temperatures but lower $\tilde{G}$ levels. Like Regime I, the~second phase is 
characterized by relatively adiabatic lattice dynamics. In~other words,  the~deformation 
landscape appears as if it is stationary for an electron, at~least for short times of $\sim$$2 
\pi/\omega_D$. This fact is further confirmed by our IPR results below. Furthermore, this 
regime is perturbative but also classical-like, meaning that the wavelength of the electron 
is shorter than the effective shortest length scale of the deformation potential. As~
thoroughly discussed in Ref.~\cite{Heller22}, the~perturbation theory pathway, 
particularly Fermi’s golden rule, is proven to be highly successful in this~phase.

The final phase (Phase III) identified by the ML-clustering is associated with highly 
nonperturbative ($\tilde{K} \lesssim 1$) electron--lattice interaction, primarily existing in 
the parameter space where electron dynamics can be considered as coherent 
($\tilde{\lambda} \lesssim 1$). Therefore, wave interference and diffraction effects are 
important because the electron wavelength is larger than the shortest length scale of the 
deformation potential. Notably, this phase begins at low temperature as $\tilde{T} 
\approx 0.5$ while extending to very high temperatures across the range of $\tilde{T}$ 
investigated.

In Phase III, as~depicted in the right column of Figure~\ref{fig:wavefunc_dynamic}, an~initial wavepacket encounters significant scattering from a strong deformation potential, initially causing diffusive behavior akin to that seen in Phase II. However, wavepacket spreading eventually ceases due to quantum interference effects, signifying an onset of localization.  Nevertheless, the~random fluctuations introduced by the motion of the lattice slowly but surely scramble the quantum interference required for the long-term confinement of the wavepacket, resulting in the transient nature of this localization (for capturing the essential aspects of this phenomenon, see, e.g.,~Ref.~\cite{Phys.Rev.Lett_118_036602_2017}). 

To achieve a more comprehensive understanding, we adopt a static potential approximation, wherein the temporal aspect of the lattice deformation field of Equation~(\ref{eq:VDcl}) is neglected, effectively frozen into its original configuration. Within~this frozen potential framework, we carry out an analysis analogous to that of the evolving deformation potential as above. This is outlined in Appendix~\ref{Appendix:Clustering_static}. Employing the same cluster classification, we categorize the data of an electron wavepacket evolution under a static deformation potential, yielding a phase diagram similar to its dynamic counterpart in Figure~\ref{fig:clusters_dynamic}, albeit with slightly sharper phase boundaries. This comparison validates treating the deformation potential as predominantly static, particularly in Phases I and II. On~the contrary, in~Phase III, the~static deformation potential results in full Anderson localization of the wavepacket that proves transient when the deformation potential undergoes morphing and undulation over time, as~further elucidated in the subsequent~analysis. 
 
\subsection{Transient~Localization}

In this section, we delve deeper into the nature of transient localization induced by lattice vibrations taking place within Phase III. At~timescales shorter than the characteristic timescale of $2\pi/\omega_D \sim 100\,\textrm{fs}$, lattice vibrations mimic a static, internal disorder field, triggering the onset of Anderson localization. Therefore, in~reference to the dynamical field where the motion of the lattice disrupts the process of Anderson localization, we are also exploring the localization behavior of a wavepacket within a frozen potential approximation. In~both cases, we quantify the level of localization by studying the time-dependent IPR, denoted as $\alpha(t)$ in Equation~\eqref{eq:IPR}, and~introducing a subsidiary product of the MSD, called the {instantaneous diffusivity}
\begin{eqnarray}
    D(t) = \frac{1}{4}\frac{d\beta(t)}{dt},
\end{eqnarray}
where $\beta(t)$ is defined in Equation~\eqref{eq:MSD}.

\begin{figure*}[ht]
\centering
\includegraphics[width=0.75\linewidth]{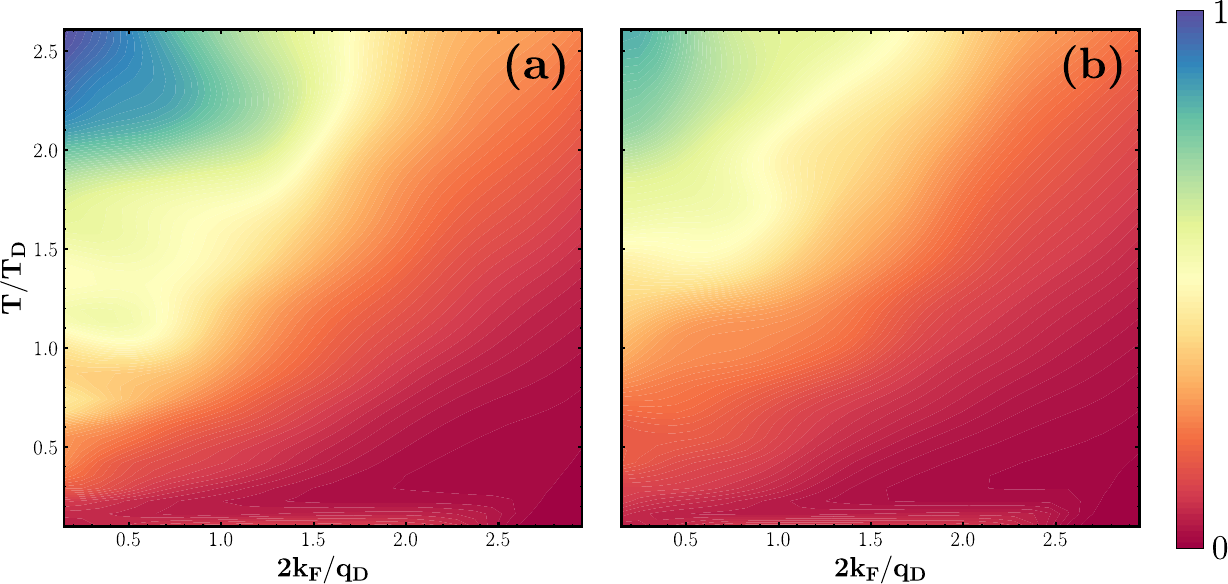}
\caption{Inverse scattering rate $1/\tau$ for electrons in LSCO in a (\textbf{a}) frozen and (\textbf{b}) dynamic deformation potential on a normalized scale. Both scenarios exhibit a high inverse scattering rate region in the upper left corner (Phase III), suggesting strong constraints on carrier mobility. This region is markedly more prominent in the frozen potential scenario, as~the lattice vibrations in the dynamic potential continuously disrupt short-lived attempts at localization. The~shape of counter lines closely resembles the dynamical phase transition lines depicted in Figure~\ref{fig:wavefunc_dynamic}.}\label{fig:scattering_rate}
\end{figure*}

We begin by examining the instantaneous diffusivity, which then determines the diffusion constant $D$ as its long-term value, i.e.,~$D = \lim_{t \to \infty} D(t)$. In~the spirit of the Einstein and Drude models, we can convert this diffusion constant into an inverse scattering rate as 
\begin{equation*}
   \frac{1}{\tau}=\frac{k_B T}{m^* D}, 
\end{equation*}
consistent with the definition used in Ref.~\cite{nature_manuscript}. Figure~\ref{fig:scattering_rate} shows the inverse scattering rate $1/\tau$ for both the cases of frozen (left panel (right panel) deformation potential. Overall, the~analysis of the scattering rate supports the ML classification underlying the phase diagram shown in Figure~\ref{fig:wavefunc_dynamic}. Both scenarios exhibit a notably high inverse scattering rate within Phase III of the phase diagram (upper right corner), indicating significant constraints on carrier mobility, as~expected in the context of Anderson localization. This effect is more prominent in the frozen potential approximation than in the case of the morphing deformation potential, underlining the fact that the dynamics of the deformation potential continuously disrupt short-lived localization attempts. Moreover, we observe that the contour lines of the inverse scattering time in Figure~\ref{fig:scattering_rate}  closely resemble the phase boundaries seen in Figure~\ref{fig:wavefunc_dynamic}.

\begin{figure*}[t!]
\centering
\includegraphics[width=\linewidth]{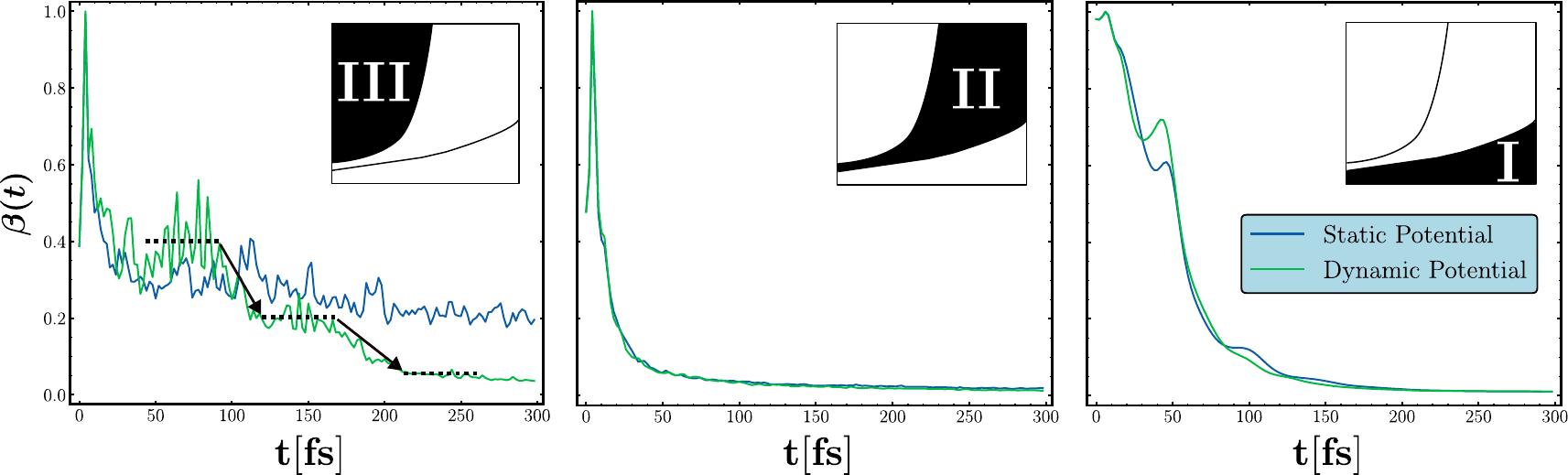}
\caption{Normalized
 inverse participation ratio $\beta(t)$ of the wavepacket as a function 
of time across three distinct dynamical regions identified in the machine learning-based 
phase diagram shown in Figure~\ref{fig:wavefunc_dynamic}. Figure contrasts the 
behaviors under static (blue line) and dynamic (green line) deformation potentials. In Phase III (left), brief localization periods (indicated by dashed lines) are broken by lattice motion (indicated by arrows), while full localization, indicated by the saturation of $\beta(t)$, occurs 
within the frozen potential approximation. In Phase~II (middle) and Phase I (right), the decay of $\beta(t)$ follows a relatively exponential trend without signs of localization as 
observed in Phase III. Notably, the~decay is faster in Phase I when compared to Phase II, 
where small oscillations linked to quantum interference are~present. \label{fig:IPR}}
\end{figure*}

A deeper insight into the emergence of transient localization is obtained by investigating 
the IPR of the wavepacket evolving over time. In~the left panel of Figure~\ref{fig:IPR}, we 
show the evolution of the IPR in Phase III for both static (blue) and dynamic (green) 
deformation potentials. In~the approximation where the deformation potential remains 
static at its initial state, the~IPR stabilizes at a certain value ($\beta \sim 0.3$) after an 
initial decrease, heralding the Anderson localization of the wavepacket. Similarly, when 
the wavepacket is subject to a dynamic deformation potential, a~form of Anderson 
localization occurs. This localized state is eventually disrupted by lattice motion, leading to 
a temporary delocalization followed by a brief relocalization before being disintegrated 
again by potential evolution. This cyclical process of plateauing IPR seen in the left panel 
of Figure~\ref{fig:IPR} epitomizes the birth and demise of the Anderson localization due 
to lattice~vibrations.

We would like to point out that the dynamics of short wavelength components of the 
deformation potential can significantly influence localization within a time window 
shorter than the characteristic time $\omega_D/2\pi$. For~example, the~left panel of 
Figure~\ref{fig:IPR} demonstrates an instance of dynamically enhanced localization: the 
random initial configuration of the potential creates valleys (mountains) that quickly move 
towards (away) the wavefunction, causing boosted localization. Vice~versa, these types of 
small potential variations in time can also lead to weaker localization compared to the 
static potential case in the first localization plateau seen in the left panel of 
Figure~\ref{fig:IPR}.

To provide a full picture, we also present the evolution of the IPR measure for Phase I and 
II in the right and left panels of Figure~\ref{fig:IPR}, respectively. Neither phase exhibits 
any signs of localization, contrasting with the behavior observed in Phase III. Moreover, 
the overlapping curves of the static and dynamic potentials further support the earlier 
assertion that the deformation potential can be effectively approximated as a static entity 
in Phase~I and II.  In Phase II, the IPR shows rapid exponential decay, quickly approaching 
the ergodic (fully delocalized) limit of $\beta\sim 0$. This behavior resembles that of a 
system with a high density of impurities characterized by Gaussian statistics. Similarly, in
Phase I, the IPR decreases towards the ergodic limit, albeit at a slower rate, displaying 
subtle oscillation. The slower, non-exponential decay can be attributed to the weak, 
refractive nature of wavepacket scattering, in~conjunction with quantum coherence and 
interference effects.

%%%%%%%%%%%%%%%%%%%%%%%%%%%%%%%%%%%%%%%%%%%%%%%%%%%%%%%%%%%%%%%%%%%%%%%

\section{Conclusion and Future~Directions}\label{Sec:conclusion}

Quantum acoustics opens up an unexplored pathway to investigating the intricate 
Fr{\"o}hlichian electron--lattice interaction inaccessible with the standard methods of 
perturbation theory. We take the concept further by treating it not just as a dual 
perspective on lattice vibrations, but~as a versatile tool in ascendance: 
a time-dependent, nonperturbative approach for electron--lattice interaction in coordinate 
space. Moreover, in the quasi-classical limit of the coherent state formalism, the quantum-acoustical way unveils the dynamics of electrons navigating through an internal 
lattice disorder field undulating and propagating in time.

In particular, we have here demonstrated the efficacy of unsupervised machine learning techniques in categorizing and analyzing the intricate aspects of electron dynamics stemming from lattice vibrations. Specifically, we have unraveled three distinct phases of behavior: refractive scattering, diffraction, and transient localization. Subsequently, we have assayed the latter phase, where the Anderson localization attempts of electron wavepackets are periodically disrupted by lattice movement, further enlightening an enigmatic phenomenon suggested to underpin the mysteries surrounding strange~metals.

Our study, supported by machine learning, explores the parameter space characterized by 
temperature and effective coupling, focusing on the paradigmatic strange metal LSCO, 
known for its two-dimensional transport behavior. However, the presented method can be 
readily extended to variations in any set of material parameters---potentially augmented 
by density functional simulations as they are not necessarily independent---and 
generalized to electron--lattice dynamics in three dimensions. Therefore, our work not 
only lays the groundwork for uncovering hidden realms in  electron--lattice interaction but 
is also a testament to designing materials with customized features by employing machine 
learning techniques from a dynamics~perspective.

%%%%%%%%%%%%%%%%%%%%%%%%%%%%%%%%%%%%%%%%%%%%%%%%%%%%%%%%%%%%%%%%%%%%%%%
\setcounter{secnumdepth}{1}
\begin{acknowledgments}

Y.Z. wants to thank the Günthard Foundation and the Swiss-European Mobility Programme (SEMP) for financial support.  J.K.-R. thanks the Oskar Huttunen Foundation for financial support. A. M. G. thanks the Harvard Quantum Initiative for financial support. The numerical simulations were performed on the Euler cluster operated by the High Performance Computing group at ETH Zurich.
\end{acknowledgments}

\section*{Appendix}
\appendix
\counterwithin{figure}{section}
\section[\appendixname~\thesection.~Clustering Analysis]{Clustering Analysis} \label{Appendix:Clustering}
For all cluster analyses in this work, the~$k$-means implementation in the {TimeSeriesKMeans} module of the tslearn package~\cite{JMLR:v21:20-091} was used. We give a brief overview of the method~here.

\subsection[\appendixname~\thesubsection.~$k$-Means]{$k$-Means}
The goal of the $k$-means algorithm, which is attributed to Lloyd~\cite{Lloyd1982k-means}, is to divide $\mathbf{X} = \{x_1, x_2, \ldots, x_N\}$ with $x_i \in \mathbb{R}^n$ into disjoint sets ({clusters}) $\mathcal{S} = \{S_1, S_2, \ldots, S_k\}$, $S_i \subseteq \mathbf{X}$, while minimizing the in-set variance. More concretely, one aims to find
\begin{eqnarray}
   \underset{\mathcal{S}}{\arg \min} \, \mathcal{L}(\mathcal{S}) &\equiv& \underset{\mathcal{S}}{\arg \min} \sum_{i=1}^{k} \sum_{x_j \in S_i} \lVert x_j - \mu_i \rVert^2 \nonumber\\ &=& \underset{\mathcal{S}}{\arg \min} \sum_{i=1}^{k} |S_i| \text{Var}\left[S_i\right], \label{Eq:K-MeansHeuristic}
\end{eqnarray}
where $\lVert \cdot \rVert$ is the Euclidean norm,  $|S_i|$ denotes the cardinality of 
$S_i$, i.e.,~the size of cluster $S_i$, and~$\mu_i$ is the cluster center (\emph{centroid}) of 
the $i$th cluster, defined as follows:
\begin{eqnarray} \label{Eq:centroids}
   \mu_i = \frac{1}{|S_i|} \sum_{x_j \in S_i} x_j
\end{eqnarray}
After some initialization scheme places the initial centroids, the~algorithm alternates between two~steps:
\begin{enumerate}
    \item {Assignment Step:} Assign all $x_i$ to their closest centroid, as~measured by the squared Euclidean distance. This defines the cluster memberships $\mathcal{S}$.
    \item {Update Step:} Update the centroids using $\mathcal{S}$ according to Equation~\eqref{Eq:centroids}.
\end{enumerate}

\begin{figure}[ht]
\includegraphics[width=\linewidth]{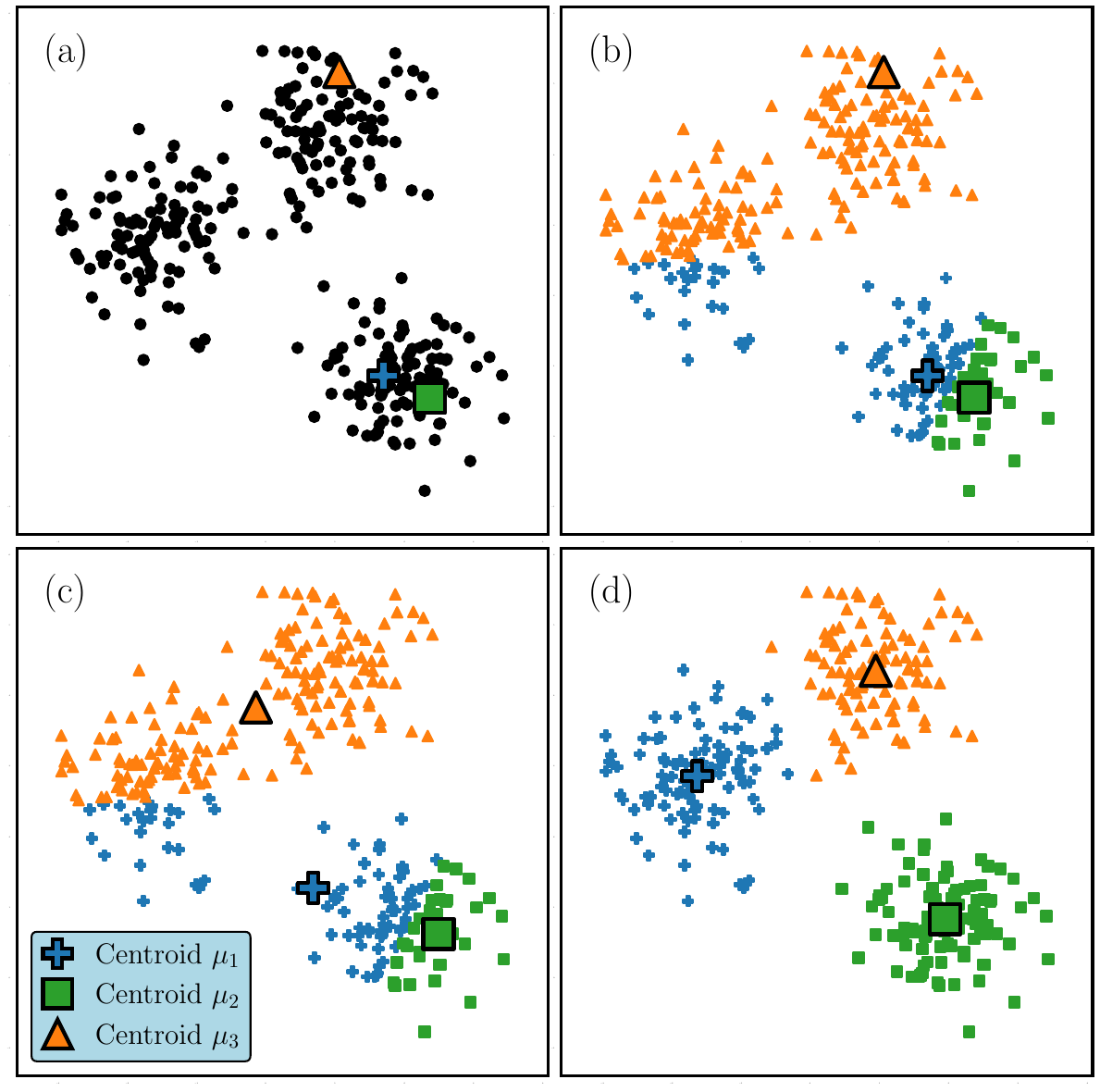}
\caption{{Toy}
 example to illustrate a run of the $k$-means algorithm with hyperparameter 
$k=3$ on example data in $\mathbb{R}^2$ (black dots). (\textbf{a})~Initial centroids are chosen by an initialization scheme (here: random 
initialization). (\textbf{b}) Assignment step of the first iteration. (\textbf{c}) Update step of 
the first iteration. (\textbf{d}) Final result. The~algorithm converges after two consequent 
iterations yield sufficiently similar clusters (here the $10$th 
iteration).\label{fig:KMeans_expainer}}
\end{figure}

{It}
 is easy to see that these steps cannot increase the in-set variance; the algorithm is 
guaranteed to converge. However, there is an important caveat in that the objective 
function $\mathcal{L}(\mathcal{S})$ defined in Equation~\eqref{Eq:K-MeansHeuristic} is 
non-convex and the algorithm can therefore converge to a local optimum that is not the 
global minimum. This problem can be mitigated by an intelligent choice of initialization 
scheme, most commonly $k${-means++}~\cite{Arthur_2007}, and~by running the 
algorithm multiple times and taking the clustering with the minimal 
$\mathcal{L}(\mathcal{S})$ post-convergence~\cite{james2013introduction}. An~example 
run of $k$-means on toy data in $\mathbb{R}^2$ is shown in 
Figure~\ref{fig:KMeans_expainer}.
\unskip
\subsection[\appendixname~\thesubsection.~Dynamic Time Warping]{Dynamic Time Warping}
\begin{figure}[ht]
\includegraphics[width=\linewidth]{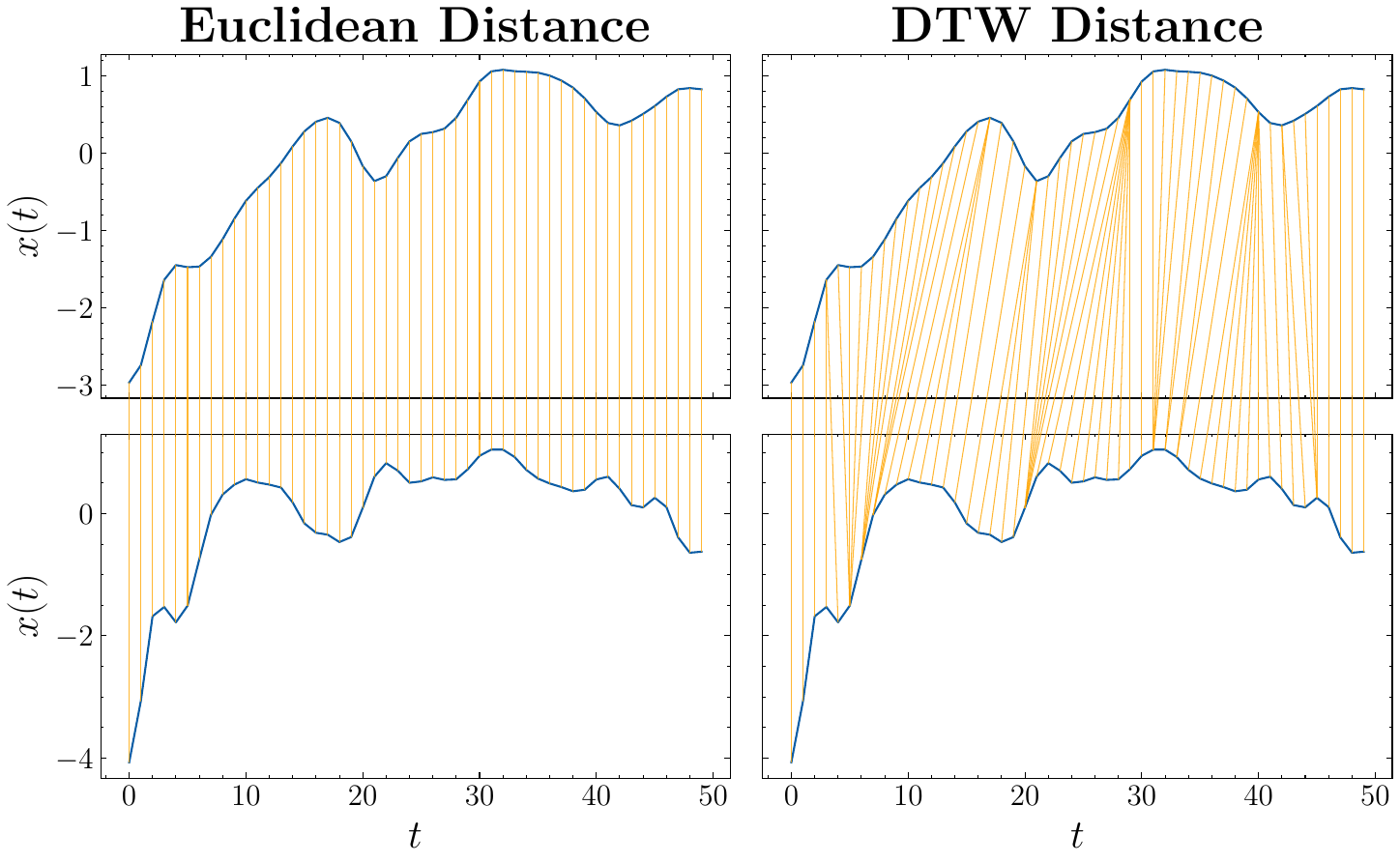}
\caption{Illustration of the difference between the Euclidean distance and Dynamic Time 
Warping (DTW) distance for two example MSD sequences from our data set. While the 
Euclidean distance compares values at corresponding time steps, DTW temporally aligns 
the sequences, providing a more robust distance measure for time series 
data.\label{fig:DTW_expainer}}
\end{figure}

When working with time series data $x(t) = (x^{t_1}, x^{t_2}, \ldots, x^{t_M})$, $x^{t_i} \in 
\mathbb{R}^d$, one could naively embed $x(t)$ in the space $\mathbb{R}^{M \times d}$ 
(here, $M = 50$ and $d=2$) equipped with the Euclidean metric and run $k$-means exactly 
as described above. However, this approach has significant limitations. It only works for 
sequences of equal length, and~more critically, the~Euclidean metric is unable to account 
for the temporal nature of the sequences. By~operating solely on elements with the same 
time indices, it ignores potential temporal misalignment between sequences. 
Consequently, even time series with similar features can result in a large metric distance if 
their phases differ. This makes it an ineffective measure for determining similarity. 
Dynamic time warping (DTW), introduced by Sakoe and Chiba~\cite{sakoe1978dynamic}, 
has been employed to overcome these challenges by identifying the most temporally 
appropriate pairs of time indices for comparison~\cite{berndt1994using}. Formally,
\begin{eqnarray}
  \text{DTW}(x(t), y(t)) = 
  \min_\pi \sqrt{\sum_{(t_i, t_j)\in\pi} \lVert x^{t_i} - y^{t_j} \rVert^2}
\end{eqnarray}
where the time index pairs $\pi = \{\pi_0, \pi_1, \ldots, \pi_K\}$ satisfy certain (boundary) conditions that we will not elaborate on for brevity (for an extensive introduction to the topic, see, e.g.,~Ref.~\cite{Müller2007}). One thus ends up with a new{,}
{warped} time path for the comparison of the time series, as~is illustrated in Figure~\ref{fig:DTW_expainer}.

DTW is not a metric, because~$\text{DTW}(x(t), y(t)) = 0$ does not imply $x(t) = y(t)$ and it 
does not obey the triangle inequality. Nevertheless, the~Fréchet mean, which generalizes 
Equation~\eqref{Eq:centroids} to any metric space, is used to compute a centroid-like 
representation, commonly referred to as {barycenter}:
\begin{eqnarray}
    \mu_i(t) = \underset{z(t)}{\arg \min} \sum_{x_j(t) \in S_i} \text{DTW}(z(t), x_j(t))^2
\end{eqnarray}
This problem is computationally NP-hard~\cite{bruning2024number}. DTW Barycenter Averaging (DBA), proposed by Petitjean~et~al.~\cite{petitjean2011global}, is a widely used algorithm to approximate the~barycenters. 

\subsection[\appendixname~\thesubsection.~Feature Scaling]{Feature Scaling}
Since $k$-means depends on distances, it is important to scale the features before performing the optimization. In~this analysis, we normalize the features to zero mean and unit variance in the time dimension:
\begin{eqnarray*}
    \tilde{x}(t) = \frac{x(t) - \mu_t}{\sigma_t},
\end{eqnarray*}
where $$\mu_t = \frac{1}{T}\int_0^T x(t) \, \text{d}t$$ and $$\sigma_t = \sqrt{\frac{1}{T} \int_0^T \left(x(t) - \mu_t\right)^2 \, \text{d}t}$$.
\subsection[\appendixname~\thesubsection.~Choosing $k$]{Choosing $k$}
\begin{figure}[ht]
\includegraphics[width=\linewidth]{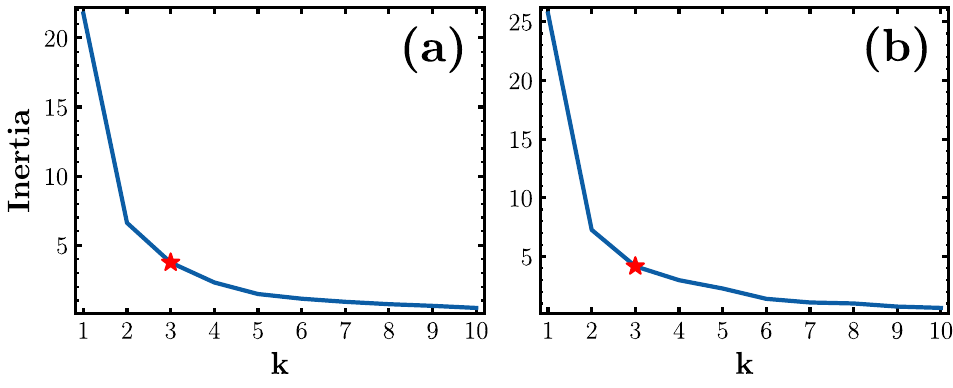}
\caption{{Elbow}
 plot depicting the sum of squared distances of samples to their nearest cluster center for varying values of $k$ for the (\textbf{a}) frozen and (\textbf{b}) dynamic lattice vibration field scenario. The~{elbow} point (indicated by a red star), where the curve shows a distinct bend, suggests the optimal number of clusters for our analysis, as~increasing the number of clusters beyond this value yields diminishing returns in terms of model~improvement.\label{fig:elbow}}
\end{figure}

The parameter $k$, i.e.,~the number of clusters, has to be set before performing the 
optimization Equation~\eqref{Eq:K-MeansHeuristic}. To~find a suitable $k$ for our analysis, we 
run the $k$-means algorithm multiple times with varying values of $k$ ranging from 1 to 
10. Figure~\ref{fig:elbow} shows a plot of the sum of squared distances of samples to their 
nearest cluster center (inertia) against different values of $k$. The~point where the curve 
bends or forms an {elbow} is considered an indicator of the natural number of 
clusters in the underlying data~\cite{james2013introduction}; additional clusters beyond 
this point do not substantially improve the model's representation of the different 
dynamical regimes of the~system.

\subsection[\appendixname~\thesubsection]{Choosing $k$}

\subsection[\appendixname~\thesubsection.~Ensemble Averaging]{Ensemble Averaging}
To mitigate the effect of statistical fluctuations of the lattice vibration field on the clustering of the electron dynamics, we perform the optimization independently on an ensemble of 10 wavefunction data sets, each generated using deformation potentials with different random initializations. The~different clustering results are then consolidated into one phase diagram by taking the consensus of the ensemble, with~the size of the dots in Figure~\ref{fig:clusters_dynamic} indicating the degree of~consensus.
\section[\appendixname~\thesection.~Clustering in the Frozen Approximation]{Clustering in the Frozen Approximation} \label{Appendix:Clustering_static}
This appendix presents the clustering analysis results using the frozen field approximation, where the deformation potential is treated as static. This simplifies the system by ignoring the temporal evolution of lattice vibrations, allowing for a direct comparison with the dynamic scenario discussed in the main~text.

Using the same $k$-means clustering technique as described in Appendix~\ref{Appendix:Clustering}, we simulate wavepacket evolution under static deformation potentials. These potentials are generated with identical parameters to those used for the dynamic scenario. We then analyze the resulting time-dependent wavefunctions using mean squared displacement (MSD) and inverse participation ratio (IPR) as~features.

\begin{figure}[ht]
\includegraphics[width=\linewidth]{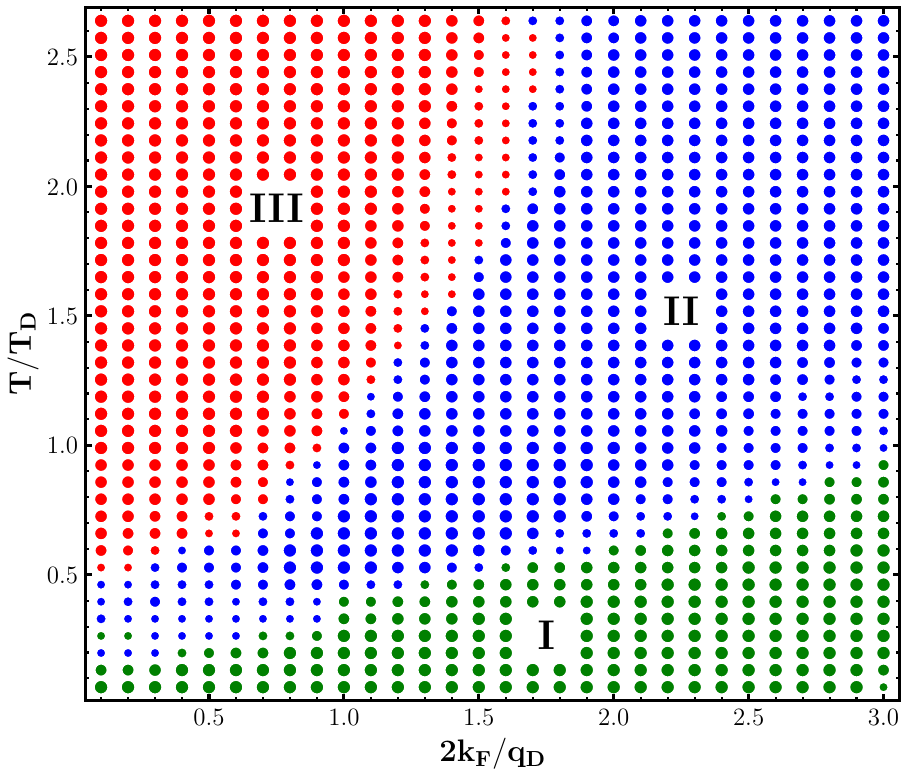}
\caption{{Phase}
 diagram of LSCO in the frozen potential field. The~phase diagram was derived using a machine learning-based clustering algorithm to analyze time series data of the wavefunction evolution within systematically varied deformation potentials. This analysis involved variations in temperature $\tilde{T}$ and effective coupling $\tilde{G}$; every point corresponds to one unique configuration. The~following three clusters were identified: (I) refractive scattering region (green), (II) diffraction behavior (blue), and (III)  Anderson localization (red). The~size of the points indicates the level of agreement across an ensemble of different wavefunction data~sets.\label{fig:clusters_static}}
\end{figure}

Figure~\ref{fig:clusters_static} shows the phase diagram for LSCO derived from clustering in the frozen approximation. As~in the dynamic case, three phases are identified. Supported by the arguments presented in the main text and especially Figure~\ref{fig:IPR}, Phases I and II are very similar to the dynamic case as the electron dynamics are highly adiabatic in these regimes. The~key difference emerges in Phase III, the~Anderson localization phase. Here, strong localization due to quantum interference in the static potential field leads to complete Anderson localization, contrasting sharply with the transient localization observed in the dynamic~case.

%%%%%%%%%%%%%%%%%%%%%%%%%%%%%%%%%%%%%%%%%%%%%%%%%%%%%%%%%%%%%%%%%%%%%%%

\end{document}